\documentstyle[preprint,aps]{revtex}

\newcommand{\beq}{\begin{equation}}
\newcommand{\eeq}{\end{equation}}
\newcommand{\beqa}{\begin{eqnarray}}
\newcommand{\eeqa}{\end{eqnarray}}
\newcommand{\klpnn}{\mbox{$K_L \to \pi^0 \nu \bar \nu$}}
\newcommand{\klpnnij}{\mbox{$K_L \to \pi^0 \nu_i \bar \nu_j$}}
\newcommand{\klpnnji}{\mbox{$K_L \to \pi^0 \nu_j \bar \nu_i$}}
\newcommand{\kspnn}{\mbox{$K_S \to \pi^0 \nu \bar \nu$}}

\newcommand{\kppnn}{\mbox{$K^+ \to \pi^+ \nu \bar \nu$}}
\newcommand{\kppnnij}{\mbox{$K^+ \to \pi^+ \nu_i \bar \nu_j$}}
\newcommand{\kppnnji}{\mbox{$K^+ \to \pi^+ \nu_j \bar \nu_i$}}
\newcommand{\kzpnn}{\mbox{$K^0 \to \pi^0 \nu \bar \nu$}}
\newcommand{\kpnn}{\mbox{$K \to \pi \nu \bar \nu$}}
\newcommand{\kpnnij}{\mbox{$K \to \pi \nu_i \bar \nu_j$}}

\def\O{{\cal O}}
\def\epsK{\varepsilon}
\def\BR{\mbox{{\rm BR}}}
\def\SM{Standard Model}
\def\sm{Standard Model}
\def\NP{New Physics}
\def\Re{\mbox{{\rm Re}}}
\def\Im{\mbox{{\rm Im}}}
\def\ris{r_{\rm is}}

\def\npb#1{Nucl.\ Phys.\ {\bf B #1}}
\def\plb#1{Phys.\ Lett.\ {\bf B #1}}
\def\prd#1{Phys.\ Rev.\ {\bf D #1}}
\def\prl#1{Phys.\ Rev.\ Lett. {\bf #1}}

\def\zpc#1{Z.~Phys.\ {\bf C #1}}

\def\ijmpa#1{Int.\ J.\ Mod.\ Phys.\ {\bf A #1}}

\begin{document}

\draft

{\tighten
\preprint{
\vbox{
      \hbox{SLAC-PUB-7380}
      \hbox{WIS-97/3/Jan-PH}
      \hbox{hep-ph/9701313}}}

\bigskip
\bigskip

\renewcommand{\thefootnote}{\fnsymbol{footnote}}

\title{\klpnn\ Beyond the \sm\ }
\author{Yuval Grossman\,$^a$ and  Yosef Nir\,$^b$}
\address{ \vbox{\vskip 0.truecm}
$^a$Stanford Linear Accelerator Center \\
        Stanford University, Stanford, CA 94309 \\
\vbox{\vskip 0.truecm}
  $^b$Department of Particle Physics \\
  Weizmann Institute of Science, Rehovot 76100, Israel}

\maketitle

\begin{abstract}%
We analyze the decay \klpnn\ in a model independent way. If lepton
flavor is conserved the final state is (to a good approximation) purely
CP even. In that case
this decay mode goes mainly through CP violating interference
between mixing and decay. Consequently, a theoretically clean
relation between the measured rate and electroweak parameters holds
in any given model. Specifically, $\Gamma(\klpnn)/\Gamma(\kppnn)=
\sin^2\theta$ (up to known isospin corrections), where $\theta$
is the relative CP violating phase between the $K-\bar K$ mixing
amplitude and the $s\to d\nu\bar\nu$ decay amplitude. The experimental
bound on $\BR(\kppnn)$ provides a model independent upper bound:
$\BR(\klpnn) < 1.1 \times 10^{-8}$. In models with lepton flavor
violation, the final state is not necessarily a CP eigenstate. Then
CP conserving contributions can dominate the decay rate.

\end{abstract}

} 

\newpage


In the \SM\ \klpnn\ is dominantly a CP violating decay \cite{Litt}.
The main contributions come from penguin and box diagrams with
an intermediate top quark and can be calculated with very little
theoretical uncertainty \cite{Buraskl,BuBu}. It then provides a clean
measurement of the Wolfenstein CP violating parameter $\eta$ or,
equivalently, of the Jarlskog measure of CP violation $J$ and, together
with \kppnn, of the angle $\beta$ of the unitarity triangle \cite{BuBu}.
The \SM\ predictions are
$\BR(\kppnn) = (9.1 \pm 3.2) \times 10^{-11}$
and $\BR(\klpnn) = (2.8 \pm 1.7) \times 10^{-11}$ \cite{Burasrev}. Such
rates are within the reach of near future experiments \cite{Burasrev}.
The \sm\ contributions to the amplitude are fourth order in the weak
coupling and proportional to small CKM matrix elements. Consequently,
this decay can be sensitive to New Physics effects \cite{CDS}.

In this paper we study the \kpnn\ decay in a model independent way.
We are mainly interested in the question of what can be learned in general
if a rate for \klpnn\ much larger than the Standard Model prediction
is observed. We find that the information from a measurement of the
rate is particularly clean and simple to interpret if lepton flavor
is conserved. In this case the \klpnn\ decay is dominated by
CP violation in the interference between mixing and decay. The
theoretical calculation of the decay rate is then free of hadronic
uncertainties and allows the clean determination of CP violating
parameters even in the presence of \NP. Knowledge of neither magnitudes
of decay amplitudes nor strong phases is required. Models with
$Z$-mediated flavor changing neutral currents serve as an example of
these points. In models with lepton flavor violation, the final
$\pi^0\nu\bar\nu$ state is not necessarily a CP eigenstate. We show that
in this case the CP conserving contributions can be significant and
even dominant. The results are still informative but more complicated to
interpret, as they depend on both CP violating and lepton flavor
violating parameters. We give an explicit example of models with
leptoquarks (or, equivalently, supersymmetry without $R$-parity).

Our notation follows refs. \cite{Breviw,BSN}.
We define the decay amplitudes $A$ and $\bar A$,
\beq \label{defA}
A=\langle\pi^0\nu\bar\nu|H|K^0\rangle, \qquad
\bar A=\langle\pi^0\nu\bar\nu|H|\bar K^0\rangle.
\eeq
If the final $\pi^0\nu\bar\nu$ is a CP eigenstate then in the CP limit
$|\bar A/A|=1$; if it is not then $\bar A$ and $A$ are not related by a
CP transformation. We further define the components of interaction
eigenstates in mass eigenstates, $p$ and $q$:
\beq \label{defqp}
K_{L,S}=p |K^0 \rangle \mp q |\bar K^0 \rangle.
\eeq
Note that $|q/p|$ is measured by the CP asymmetry in $K_L\to\pi\ell\nu$
and is very close to unity: $1-|q/p|=2 \Re\ \epsK$. Finally, we define
a quantity $\lambda$,
\beq \label{deflam}
\lambda \equiv {q \over p} {\bar A \over A}.
\eeq
The decay amplitudes of $K_L$ and $K_S$ into a final $\pi^0\nu\bar\nu$
state are then
\beq \label{AKLS}
\langle\pi^0\nu\bar\nu|H|K_{L,S}\rangle = p A \mp q \bar A,
\eeq
and the ratio between the corresponding decay rates is
\beq \label{ratrat}
{\Gamma(\klpnn) \over \Gamma(\kspnn)} =
{1 + |\lambda|^2 - 2\Re\lambda\over 1 + |\lambda|^2 + 2\Re\lambda}.
\eeq

We first assume that the final state is purely CP even. This is the
case to a good approximation when lepton flavor is conserved. In general,
a three body final state does not have a definite CP parity. However, for
purely left-handed neutrinos (which is presumably the case if neutrinos
are massless), the lowest dimension term in the effective Hamiltonian
relevant to \klpnn\ decay is $K(\partial_\mu\pi)(\overline\nu_{iL}
\gamma^\mu \nu_{iL})$. Using the CP transformation
properties of the leptonic current, we find that this interaction
`forces' the $\nu_i\bar\nu_i$ system into a state of
well-defined CP, namely CP even. As far as Lorentz and CP transformation
properties are concerned, we can then think of the final $\pi\nu\bar\nu$
state as a two body $\pi Z^*$ state which, when produced by $K_L$ decay
(namely, carrying total angular momentum $J=0$), is CP even
\cite{BakGla,AmbIsi}. Higher dimension operators can induce CP conserving
contributions. For example, $K(\partial_\nu\partial_\mu\pi)
(\overline\nu_{iL}\gamma^\mu{\buildrel{\leftrightarrow}\over
{\partial^\nu}}\nu_{iL})$ will lead
to an amplitude that is proportional to $p_\pi\cdot(p_\nu-p_{\bar\nu})$
and, consequently, to a CP odd final state. However, these contributions
are $\O(m_K^2/m_W^2)\sim10^{-4}$ compared to the leading CP violating
ones and can be safely neglected. (In the \SM\ this operator arises from
the box diagram when external momenta are not neglected.) With massive
neutrinos, new CP conserving operators arise,
{\it e.g.} $K\pi (\overline{\nu_i}\nu_i)$.
The final state is now equivalent (in the
Lorentz and CP properties) to a two body $\pi H^*$ state (where $H$ is
a scalar), which is CP odd. However, this amplitude is proportional to
the neutrino mass and again negligible. We conclude then that,
for any model where
lepton flavor is conserved, the CP conserving transition amplitude
for \klpnn\ is highly suppressed and can be neglected.

If the final state $\pi^0\nu\bar\nu$ is CP even, then \klpnn\ vanishes in
the CP limit. This can be seen directly from eq. (\ref{ratrat}):
if CP is a good symmetry then $|q/p|=1$, $|\bar A/A|=1$ and $\lambda=1$.
With CP violation we can still neglect CP violation in the mixing
($|q/p|\neq1$) and in the decay ($|\bar A/A|\neq1$). As mentioned above,
the deviation of $|q/p|$ from unity is experimentally measured
and is $\O(10^{-3})$. The deviation of $|\bar A/A|$ from unity is
expected to be even smaller: such an effect requires contributions to the
decay amplitude which differ in both strong and weak phases
\cite{Breviw}. While in the presence of \NP\ we could easily have more
than a single weak phase involved, we do not expect
the various amplitudes to differ in their strong phases. An absorptive
phase comes from light intermediate states. In the language of
quark subprocesses, only an intermediate up quark could contribute.
But there is a hard GIM suppression that makes these contributions
negligibly small \cite{ReSe,HaLi,LuWi,GeHsLi,Fajfer,BuBu}.
Therefore, it is safe to assume that $|\lambda|=1$ to $\O(10^{-3})$
accuracy. The leading CP violating effect is then $\Im\lambda\neq0$,
namely interference between mixing and decay. This puts the ratio of
decay rates (\ref{ratrat}) in the same class as CP asymmetries in various
$B$ decays to final CP eigenstates, e.g. $B\to\psi K_S$, where a very
clean theoretical analysis is possible \cite{Breviw}.

As a result of this cleanliness, the CP violating phase can be extracted
almost without any hadronic uncertainty, even if this phase comes from
\NP. Specifically, defining $\theta$ to be the relative phase between the
$K-\bar K$ mixing amplitude and the $s\to d\nu\bar\nu$ decay amplitude,
namely $\lambda=e^{2i\theta}$, we get from eq. (\ref{ratrat})
\beq \label{genrat}
{\Gamma(\klpnn) \over \Gamma(\kspnn)} =
{1 - \cos 2 \theta \over 1 + \cos 2 \theta} = \tan^2 \theta.
\eeq
This ratio measures $\theta$ without any information about the magnitude
of the decay amplitudes. In reality it will be impossible to
measure $\Gamma(\kspnn)$. We can use the isospin symmetry relation,
$A(\kzpnn)/A(\kppnn)=1/\sqrt{2}$,
to replace the denominator by the charged kaon decay mode:
\beq \label{ratis}
a_{CP} \equiv \ris {\Gamma(\klpnn) \over \Gamma(\kppnn)} =
{1 - \cos 2 \theta \over 2} = \sin^2 \theta,
\eeq
where $\ris = 0.954$ is the isospin breaking factor \cite{MaPa}.
The ratio (\ref{ratis}) may be experimentally measurable, as the relevant
branching ratios are $\O(10^{-10})$ in the \SM\ and even larger in some
of its extensions. It will provide us with a very clean measurement of
the CP violating phase $\theta$ which has a clear interpretation in any
given model.

In the \SM, the penguin and box diagrams mediating the
$s \to d \nu \bar \nu$ transition get contributions from
top and charm quarks in the loop. The charm diagrams carry the same
phase as the mixing amplitude, arg($V_{cd}V_{cs}^*$). The top diagrams
depend on arg($V_{td}V_{ts}^*$), so that their phase difference from
the mixing amplitude is the angle $\beta$ of the unitarity triangle.
Had the top contribution dominated both \klpnn\ and \kppnn, we would
have $\theta=\beta$. However, while the charm contribution to \klpnn\
is negligible, it is comparable to the top contribution to \kppnn.
Then we cannot directly relate the experimentally-derived $\theta$
of eq. (\ref{ratis}) to the model parameter $\beta$, and a calculation of
the charm and top amplitudes is also needed \cite{BuBu}. With \NP,
the magnitude of the decay amplitude is generally not known.
The ratio (\ref{ratis}) is most useful if both \klpnn\ and \kppnn\ are
dominated by the same combination of mixing angles. The phase of this
combination is then directly identified with $\theta$, and we need not
know any other of the new parameters.

Eq. (\ref{ratis}) allows us to set an upper bound on $\BR(\klpnn)$. Using
$\sin^2 \theta \le 1$ and $\tau_{K_L}/\tau_{K^+}=4.17$, we have
\beq \label{isorel}
\BR(\klpnn) < 4.4 \times \BR(\kppnn).
\eeq
Using the $90\%\,$CL experimental upper bound \cite{Adler}
\beq \label{expup}
\BR(\kppnn) < 2.4 \times 10^{-9},
\eeq
we get
\beq \label{expupl}
\BR(\klpnn) < 1.1 \times 10^{-8}.
\eeq
Actually, eq. (\ref{isorel}) assumes only isospin relations and does
not even require that the final state is CP even. Therefore, the bound
(\ref{expupl}) is model independent.
This bound is much stronger than the direct experimental upper bound
\cite{Weaver} $\BR(\klpnn) < 5.8 \times 10^{-5}$.

New Physics can modify both the mixing and the decay amplitudes.
The contribution to the mixing can be of the same order as the \SM\ one.
However, $\epsK=\O(10^{-3})$ implies that any such new contribution
to the mixing amplitude carries the same phase as the \SM\ one
(to $\O(10^{-3})$). On the other hand, the upper bound (\ref{expup})
which is about 30 times larger than the \SM\ prediction \cite{BuBu}\
allows New Physics to dominate the decay amplitude (with an arbitrary
phase). We conclude that the only relevant new contribution to $a_{CP}$
can come from the decay amplitude. This is in contrast to the $B$ system
where we expect significant effects of New Physics mainly in the mixing
amplitude (see e.g. \cite{NPBexa}).

We now give an explicit example of a New Physics model with potentially
large effects on \klpnn. We consider a model with extra quarks in
vector-like representations of the standard Model gauge group,
\beq \label{extra}
d_4(3,1)_{-1/3}  \qquad + \qquad \bar d_4(\bar 3,1)_{+1/3}.
\eeq
Such (three pairs of) quark representations appear, for example, in GUTs
with an $E_6$ gauge group. It is well known that the presence of new
heavy fermions with non-canonical $SU(2)$ transformations (left-handed
singlets and/or right-handed doublets) mixed with the standard leptons
and quarks would give rise to tree level flavor changing neutral currents
in $Z$ interactions \cite{fit}. Moreover, these flavor changing
$Z$ couplings can be CP violating \cite{NiSiZ}.
The flavor changing part of the couplings reads
\beq \label{mixcurrent}
{\cal L}_{\rm FCNC}^Z = {g\over 2\cos\theta_W}\, \sum_{i\neq j} \, \left[
\bar d_{L}^i\, U_{ij}\, \gamma^\mu\, d_{L}^j \right] Z_\mu \,.
\eeq
As the flavor changing couplings are very small, the flavor diagonal
$Z$ couplings are still very close to their \sm\ values. Assuming that
the $Z$-mediated tree diagram dominates \kpnn, we get \cite{NiSiZ,GLN}
\beq \label{Zkpnn}
{\Gamma(\kppnn)\over\Gamma(K^+ \to \pi^0 e^+ \nu )}=\ris^+{1 \over 4}
{|U_{ds}|^2\over|V_{us}|^2}\,, \qquad
{\Gamma(\klpnn)\over\Gamma(K^+ \to \pi^0 e^+ \nu )}=\ris^0{1 \over 4}
{|\Im U_{ds}|^2\over|V_{us}|^2}\,.
\eeq
Here $\ris^0=0.944$ and $\ris^+=0.901$ are the isospin breaking
corrections \cite{MaPa}  (so that $\ris=\ris^+/\ris^0$).
The ratio (\ref{ratis}) measures, in this case,
$\sin\theta=\Im U_{ds}/|U_{ds}|$.

We now show that the experimental bounds on the model parameters indeed
still allow large effects in \kpnn.
{}From $K_L \to \mu^+ \mu^-$  we get \cite{NiSiZ,SilvZ} (taking into
account uncertainties from long distance contributions \cite{Ko}),
\beq
|\Re(U_{ds})| \lesssim 2 \times 10^{-5}.
\eeq
{}From \kppnn\ we get (see (\ref{Zkpnn}) and (\ref{expup}))
\beq
|U_{ds}| \le 1.0 \times 10^{-4}.
\eeq
The measurement of $\epsK$ implies \cite{NiSiZ,SilvZ}
\beq
|\Re(U_{ds}) \, \Im(U_{ds})| \lesssim 1.3 \times 10^{-9}.
\eeq
Then indeed a strong enhancement of the \kpnn\ rates is possible.
If $|\Re(U_{ds})|$ and $|\Im(U_{ds})|$ are close to their upper bounds,
the branching ratios $\BR(\kppnn)$ and $\BR(\klpnn)$
are $O(10^{-9})$ and $a_{CP}$ of Eq. (\ref{ratis}) is $\O(1)$.
The measurement of $\BR(\kppnn)$ determines $|U_{ds}|$, and the
additional measurement of $\BR(\klpnn)$ determines $\arg(U_{ds})$.

Before turning to the investigation of models with lepton
flavor violation, we would like to clarify one more point.
It is often stated that a measurement of $\BR(\klpnn)\ge
\O(10^{-11})$ will provide a manifestation of direct CP violation.
This statement is somewhat confusing because,
as explained above, \klpnn\ at this level
is a manifestation of interference between mixing and decay,
$\Im\lambda\ne0$, and not of what is usually called direct
CP violation, namely $|\bar A/A|\neq1$. Furthermore, CP violation
in the interference of mixing and decay has already been
observed in $\Im(\varepsilon)\ne0$ (see discussion in
\cite{Breviw}). What is then meant by the above statement
is the following: the measurement of $\Im(\varepsilon)=\O(10^{-3})$
together with a measurement of $\BR(\klpnn)\ge\O(10^{-11})$ will
show that CP violation cannot be confined to $\Delta s=2$
processes (mixing) but necessarily affects $\Delta s=1$ processes
(decays) as well. More specifically, while one of the two
ratios $A(K\to\pi\pi)/A(\bar K\to\pi\pi)$ and
$A(K \to \pi^0 \nu \bar \nu)/A(\bar K \to \pi^0 \nu \bar \nu)$
can always be chosen real by convention, it will be impossible to do
so for both \cite{Breviw}. This will exclude those superweak scenarios
where CP violation appears in the mixing only.

We next explain how, in the presence of lepton flavor violating
new physics, $\Gamma(\klpnn)\neq0$ is allowed even if CP is conserved.
The crucial point is that the final state in \klpnn\ is not necessarily
a CP eigenstate anymore. Specifically, $\klpnnij$ with $i \neq j$ is
allowed. Then, $A$ and $\bar A$ of eq. (\ref{defA}) are no longer related
by a CP transformation, and we may have
\beq \label{lmnon}
\left|{\bar A_{ij} \over A_{ij}}\right| \equiv
\left|{A(\bar K^0 \to \pi \nu_i \bar \nu_j)
\over A(K^0 \to \pi \nu_i \bar \nu_j)}\right| \ne 1 \quad
\Longrightarrow \quad
|\lambda_{ij}|=\left|{q\over p}{\bar A_{ij}\over A_{ij}}\right| \ne 1,
\eeq
and the rate $\Gamma(\klpnnij)\propto(1+|\lambda_{ij}|^2
-2\Re\lambda_{ij})$ does not vanish even in the CP limit.

To gain further insight into the consequences of (\ref{lmnon}),
we note that the vanishing of strong phases implies a relation
between the transition amplitudes into $\pi^0\nu_i\bar\nu_j$ and
$\pi^0\nu_j\bar\nu_i$:
\beq \label{nodelta}
A_{ij}=\bar A_{ji}^*,\ \ \ \ \bar A_{ij}=A_{ji}^*.
\eeq
Eq. (\ref{nodelta}) together with $|q/p|=1$ give
\beq \label{lamrel}
\lambda_{ij}=(\lambda_{ji}^{-1})^*
\eeq
and $\Gamma(\klpnnij)=\Gamma(\klpnnji)$.
Recalling the isospin relations,
\beq
A(\kppnnij)=\sqrt2 A_{ij},\qquad A(\kppnnji)=\sqrt2 A_{ji},
\eeq
we find
\beq \label{ratisim}
a_{ij}\equiv \ris\, {\Gamma(\klpnnij)+\Gamma(\klpnnji)
\over\Gamma(\kppnnij)+\Gamma(\kppnnji)} =
{|1 - \lambda_{ij}|^2 \over 2( 1 + |\lambda_{ij}|^2) }.
\eeq
A few comments are in order with regard to eq. (\ref{ratisim}):
\begin{enumerate}
\item This ratio is always smaller than unity so,
as argued above, the bound (\ref{expupl}) applies also to this case.
\item Things are particularly simple if there is only a single pair
of indices $(ij)$ for which $|1-\lambda_{ij}|=\O(1)$. Then eq.
(\ref{ratisim}) gives the ratio of total rates,
$a\equiv\ris\, {\Gamma(\klpnn)\over\Gamma(\kppnn)}=a_{ij}$.
\item This ratio is invariant under $\lambda_{ij}\to
(\lambda_{ij}^{-1})^*$, as it should.
\item In the CP limit, $\lambda_{ij}$ is real and
$a_{ij}={(1 - \lambda_{ij})^2 \over 2( 1 + \lambda_{ij}^2) }$.
Note, however, that for final states that
are not CP eigenstates, the $\lambda$'s are real only if both the weak
and the strong phases vanish \cite{BSN}. This is in contrast to final
CP eigenstates for which $\lambda$ is always real in the CP limit.
\end{enumerate}

As an explicit example of lepton flavor violation we consider
a model with light leptoquarks (LQ). (This example is of particular
interest in the framework of SUSY models without $R$-parity where the
$\lambda^\prime L Q \bar d$ terms in the superpotential give the same
effects, with the $\tilde{d}$ squark playing the role of the
leptoquark.) An iso-singlet scalar leptoquark, $S_0$, couples to
neutrinos and down quarks \cite{davidson}:
\beq \label{LeffLQ}
{\cal L}_{\rm LQ} = - h_{iq}\, \bar q^c_L\, \nu_L^i\, S_0 + {\rm h.c.},
\eeq
with $i=e,\mu,\tau$ and $q=d,s,b$. Such couplings contribute to \kpnn\
through tree level LQ exchange:
\beq
A(\kpnnij) \propto {h_{is} h_{jd}^* \over m_{S_0}^2}.
\eeq
The strongest bounds on $|h_{is} h_{jd}^*|$ come from the bound
on $\BR(\kppnn)$ (Eq. (\ref{expup})) \cite{davidson}, so obviously
LQ exchange can dominate \kpnn. Neglecting the \SM\ contribution we get
\beq \label{lmnonexp}
\lambda_{ij} = {q \over p}
{h_{i s} h_{j d}^* \over h_{i d} h_{j s}^*}.
\eeq
If there is no fine-tuning we expect $1-|\lambda_{ij}|= \O(1)$
for $i\ne j$ ($|\lambda_{ii}|=1$ follows directly from (\ref{lmnonexp})).
We learn that, in this scenario, the CP conserving effect in the
$i\neq j$ channels is expected to be the same order of, or even dominate
over, the CP violating one. For example, assuming hierarchical flavor
structure (namely, $h_{iq}$ is smaller for lighter generations) and
CP symmetry (namely, $h_{iq}$ is real), we find that \klpnn\ has only CP
conserving contributions, and (barring a fine-tuned relation between
$h_{\mu d}h_{\tau s}^*$ and $h_{\tau d}h_{\mu s}^*$) dominated by
$\pi^0\nu_\mu\bar\nu_\tau$ and $\pi^0\nu_\tau\bar\nu_\mu$ final states.
Note that under the same assumptions $K^+\to\pi^+\nu_\tau\bar\nu_\tau$ is
the dominant charged decay mode and the ratio of total rates is small,
$a\ll1$. If, however, either $h_{\tau s}$ or $h_{\tau d}$ is small
(that could be a result of the interplay between horizontal symmetries
and holomorphy \cite{Hori}), then $a=O(1)$ even without CP violation.

Let us summarize our main points. In models with lepton flavor
conservation, $\BR(\klpnn) \ne 0$ signifies CP violation. More precisely,
it is a manifestation of CP violation in the interference between
mixing and decay, which allows a theoretically clean analysis.
The ratio $\BR(\klpnn)/\BR(\kppnn)$  (see Eq. (\ref{ratis}))
provides a clean measurement of a CP violating phase.
This phase can be either the \sm\ phase or one coming from \NP\
(or a combination of the two). The same ratio gives a model independent
bound on $\BR(\klpnn)$ (see Eq. (\ref{expupl})).
In general \klpnn\ can also have CP conserving contributions. These
contributions are negligible in the \SM\ and expected to be very small
in all its extensions with lepton flavor conservation. In models with
lepton flavor violation, however, CP conserving contributions can be
large, and even dominate the decay rate. A measurement of $\BR(\klpnn)$
is then guaranteed to provide us with valuable information. It will
either give a new clean measurement of CP violation,
or indicate lepton flavor violation.

\newpage

\acknowledgements
We thank Andrzej Buras for asking the questions that led to this work.
We thank Francesca Borzumati, Gerhard Buchalla, Gino Isidori,
Riccardo Rattazzi, Adam Schwimmer and Mihir Worah
for helpful discussions.
Y.G. is supported by the Department of Energy under contract
DE-AC03-76SF00515. Y.N. is supported in part by the United States --
Israel Binational Science Foundation (BSF), by the Israel Science
Foundation, and by the Minerva Foundation (Munich).

{\tighten

}


\begin{references}

\bibitem{Litt}
{L.S. Littenberg, \prd{39} (1989) 3322.}

\bibitem{Buraskl}
{G. Buchalla and A.J. Buras, \npb{400} (1993) 225;
A.J. Buras, \plb{333} (1994) 476.}

\bibitem{BuBu}
{G. Buchalla and A.J. Buras, \prd{54} (1996) 6782.}

\bibitem{Burasrev}
{A.J. Buras, hep-ph/9610461.}

\bibitem{CDS}
{G. Belanger, C.Q. Geng, and P. Turcotte, \prd{46} (1992) 2950;
C.E. Carlson, G.D. Dorata and M. Sher, \prd{54} (1996) 4393.}

\bibitem{Breviw}
For a review see, e.g.
Y. Nir, Lectures presented in the 20th SLAC Summer Institute,
SLAC-PUB-5874 (1992);
Y. Nir and H.R. Quinn, Ann. Rev. Nucl. Part. Sci. {\bf 42} (1992) 211.

\bibitem{BSN}
{G. Blaylock, A. Seiden and Y. Nir, \plb{355} (1995) 555.}


\bibitem{BakGla}
M. Baker and S.L. Glashow, Nuo. Cim. {\bf 25} (1962) 857.

\bibitem{AmbIsi}
For a review, see, e.g. G. D'Ambrosio and G. Isidori, hep-ph/9611284.

\bibitem{ReSe}
{D. Rein and L.M. Sehgal, \prd{39} (1989) 3325.}

\bibitem{HaLi}
{J.S. Hagelin and L.S. Littenberg,
 Prog. Part. Nucl. Phys. {\bf 23} (1989) 1.}

\bibitem{LuWi}
{M. Lu and M. Wise, \plb{324} (1994) 461.}

\bibitem{GeHsLi}
{C.Q. Geng, I.J. Hsu and Y.C. Lin, \plb{355} (1995) 569.}

\bibitem{Fajfer}
{S. Fajfer, hep-ph/9602322.}

\bibitem{MaPa}
{W. Marciano and Z. Parsa, \prd{53} (1996) 1.}

\bibitem{Adler}
{S. Adler {\it et al.}, BNL 787 Collaboration, \prl{76} (1996) 1421.}

\bibitem{Weaver}
{M. Weaver {\it et al.}, E799 Collaboration, \prl{72} (1994) 3758.}

\bibitem{NPBexa}
Y. Nir and D. Silverman, \npb{345} (1990) 301;
C. Dib, D. London and Y. Nir, \ijmpa{6} (1991) 1253;
T. Goto, N. Kitazawa, Y. Okada and M. Tanaka \prd{53} (1996) 6662;
N.G. Deshpande, B. Dutta and S. Oh, hep-ph/9608231;
M. Gronau and D. London, hep-ph/9608430;
J.P. Silva and L. Wolfenstein, hep-ph/9610208;
Y. Grossman and M.P. Worah, hep-ph/9612269,


\bibitem{fit}
See, for example,
B.A. Campbell {\it et al.}, \ijmpa{2} (1987) 831;
P. Langacker and D. London, \prd{38} (1988) 886; \prd{38} (1988) 907;
M. Shin, M. Bander and S. Silverman, \plb{219} (1989) 381;
E. Nardi, E. Roulet and D. Tommasini, \npb{386} (1992) 239.

\bibitem{NiSiZ}
{Y. Nir and D. Silverman, \prd{42} (1990) 1477.}

\bibitem{GLN}
Y. Grossman, Z. Ligeti and E. Nardi, \npb{369} (1996) 369.

\bibitem{SilvZ}
{D. Silverman, \prd{45} (1992) 1800.}

\bibitem{Ko}
{P. Ko, \prd{45} (1992) 174;
G. Buchalla and A.J. Buras, \npb{412} (1994) 106.}

\bibitem{davidson}
{M. Leurer, \prl{71} (1993) 1324;
S. Davidson, D. Bailey and B.A. Campbell, \zpc{61} (1994) 613.}

\bibitem{Hori}
See {\it e.g.} Y. Grossman and Y. Nir, \npb{448} (1995) 30.

\end{references}
\end{document}